\documentstyle[12pt]{article}   

\input epsf.tex

\newcommand{\be}{\begin{equation}}
\newcommand{\ee}{\end{equation}}
\newcommand{\bea}{\begin{eqnarray}}
\newcommand{\eea}{\end{eqnarray}}
\newcommand{\req}[1]{~(\ref{#1})}

\newcommand{\gs}{\mbox{$g_s$}}            
\newcommand{\ap}{\mbox{$\alpha^\prime$}}  

\newcommand{\npb}[3]{Nucl.~Phys. {\bf B#1} (#2) #3}
\newcommand{\plb}[3]{Phys.~Lett. {\bf #1B} (#2) #3}

\newcommand{\hepth}[1]
      {\href{http://xxx.lanl.gov/abs/hep-th/#1}{{\tt hep-th/#1}}}

\def\p{\partial}
\newcommand{\tr}{\mathop{\rm Tr}}
\def\brac#1{\langle #1 \rangle}
\def\href#1#2{#2}

\newcommand{\trFsq}{{\mathcal O}_{F^{2}}}

\newcommand{\cO}{{\mathcal{O}}}
\newcommand{\cN}{{\mathcal{N}}}
\newcommand{\cZ}{{\mathcal{Z}}}
\newcommand{\bS}{{\mathbf{S}}}

\newcommand{\vY}{\mbox{$\vec{Y}$}}          
\newcommand{\vX}{\mbox{$\vec{X}$}}
\newcommand{\vx}{\mbox{$\vec{x}$}}
\newcommand{\vy}{\mbox{$\vec{y}$}}
\newcommand{\tphi}{\mbox{$\tilde{\phi}$}}

\begin{document}

\begin{titlepage}

\begin{flushright}
PUPT-1870\\
hep-th/9906153
\end{flushright}

\vspace{1cm}

\begin{center}
{\huge Undulating Strings and Gauge Theory Waves}
\end{center}
\vspace{5mm}

\begin{center}
{\large Curtis G.\ Callan,
Jr.\footnote{callan@feynman.princeton.edu} and
Alberto G\"uijosa\footnote{aguijosa@princeton.edu}} \\
\vspace{3mm}
Joseph Henry Laboratories\\
Princeton University\\
Princeton, New Jersey 08544\\
\end{center}

\vspace{5mm}

\begin{center}
{\large Abstract}
\end{center}
\noindent

We study some dynamical aspects of the correspondence
between strings in $AdS$ space and external heavy quarks in $\cN=4$ SYM.
Specifically, by examining waves propagating on such strings,
we make some plausible (and some surprising) inferences about the
time-dependent fields produced by oscillating quarks in the
strongly-coupled gauge theory.
We point out a puzzle regarding energy conservation in the SYM theory.
In addition, we perform a similar
analysis of the gauge fields produced by a
baryon (represented as a D5-brane
with string-like extension in $AdS$ space) and compare and contrast
with the gauge fields produced by a quark-antiquark pair (represented
as a string looping through $AdS$ space).

\vfill
\begin{flushleft}
June 1999
\end{flushleft}
\end{titlepage}
\newpage

\section{Introduction}

In the context of Maldacena's correspondence between gauge theories and
gravity \cite{jthroat}, external charges in the gauge theory
are dual to macroscopic strings in anti-de Sitter ($AdS$)
space whose endpoints lie on the boundary.
This identification stems from the general
role of strings connecting
parallel branes as W-bosons of the corresponding
spontaneously broken worldvolume theory \cite{wittenbound},
and can be confirmed within the $AdS$/CFT setting
by computing the energy of such strings \cite{reyee,juanwilson}.

For concreteness, we will restrict attention to the duality between
$D=3+1$ $\cN=4$ $SU(N)$ super-Yang-Mills (SYM)
and Type IIB string theory on $AdS_{5}\times\bS^{5}$.
A solitary static quark (transforming in the fundamental
of $SU(N)$) corresponds to a Type IIB
string which extends solely in the radial direction;
a string of opposite orientation represents an antiquark
(transforming in the anti-fundamental of $SU(N)$).
The GKPW recipe for extracting gauge theory
expectation
values from the bulk action \cite{gkpads,wittenholo}
makes it possible to verify directly that a
radial string gives rise to the
correct point charge field configuration \cite{dkk}.
We note in passing that
expectation values due to string probes in the
bulk of $AdS$ (with no endpoints on the boundary)
have also been computed \cite{bdhm,vijay2,dkk}.

A quark-antiquark pair in the gauge theory is naturally identified
with a string with
both of its endpoints on the boundary. Expectation values of Wilson
loops can thus be deduced from the bulk theory by evaluating
the area of a string worldsheet which is bounded
by the loop \cite{reyee,juanwilson}.
The result of such a calculation encodes in particular
the quark-antiquark potential (see \cite{dornotto} for a
review of results on Wilson loops obtained
from the bulk-boundary correspondence).

A defining property of a string is its ability to undulate. The
identification of strings and charges raises an obvious question:
what is the gauge theory interpretation of string
oscillations? This is the issue we will address in what follows.
The main tool at our disposal is again the GKPW calculational
prescription \cite{gkpads,wittenholo}.
A string is a source for the supergravity fields, so an oscillating
string generates fluctuating fields in the bulk of $AdS$ space.
The correspondence then translates the fluctuating supergravity
fields on the boundary into the time-dependent SYM expectation
values associated with an oscillating charge.
The analysis thus establishes a correspondence between string
oscillations and gauge theory waves (including, one would hope, the
usual $r^{-1}$ radiation fields produced by an accelerated charge).

In Section \ref{fluctsec} we will fill in the details of the
procedure outlined in the previous paragraph. To
understand the basic ideas it will suffice to concentrate on waves
of the dilaton field, which is known to couple to the operator
\be \label{o}
\cO_{F^{2}}={1\over 4 g_{YM}^{2}}
\tr \left\{
   F^{2}+[X_{I},X_{J}] [X^{I},X^{J}]
  +\mbox{fermions}
\right\}
\ee
in the boundary theory \cite{igor,iwm}.
There is much to be learned by studying
waves of other supergravity fields, especially the graviton, but we
will leave this more difficult exercise for another paper.
In the above equation
$X^{I}$, $I=1,\ldots,6$,
denote the scalar fields
of the $\cN=4$ SYM theory (living in the vector of $SO(6)$).

The simple case of an oscillating
straight radial string will be worked out in Section \ref{straightsec}.
In Section \ref{bentsec} we  will then extend the analysis
to the more intricate case of a `bent' string, and discuss some
interesting features of the fields of a
quark-antiquark pair.  We amplify the discussion on the implications
of our results for the SYM theory in Section \ref{implisec}, where
we point out a puzzle regarding energy conservation in the gauge
theory.
In Section \ref{barsec} we apply
the same methods to obtain the gauge field profile due to
a baryon (represented as a D5-brane appropriately wrapped in $AdS$
space) and compare with the quark-antiquark case. A final section
consists of a brief summary of our conclusions.
Some aspects of
string oscillations and SYM waves have been examined before in
\cite{salamand,reyee,das,bakrey} and we have attempted to go beyond these
efforts in ways about which we will comment as appropriate.

\section{String Oscillations Make SYM Waves}
\label{fluctsec}

\begin{figure}[htb]
\centerline{\epsfxsize=9cm
\epsfbox{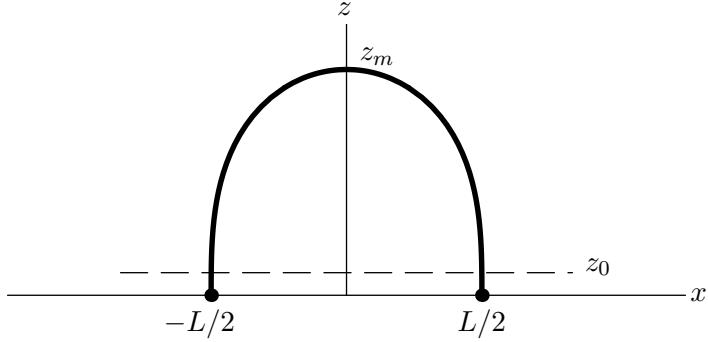}}
\caption{\small A Nambu-Goto string lying along a geodesic, with its
two endpoints on the boundary.}
\label{bentstring}
 \begin{picture}(0,0)
 \put(52,19){\small $-L/2$}
 \put(91,19){\small $L/2$}
 \put(122,23){\small $x$}
 \put(108,27){\small $z_{0}$}
 \put(78,55){\small $z_{m}$}
 \put(75,61){\small $z$}
 \end{picture}
\end{figure}

We describe the dynamics of a fundamental string through the
Nambu-Goto action
\be \label{ng}
S_{F}= -{1\over 2\pi\ap} \int d^{2}\sigma \sqrt{-g},
\ee
where $g$ is the induced metric on the string worldsheet. We work in
Poincar\'e coordinates for $AdS_{5}$,
with the metric
\be \label{adsfiveagain}
ds^2 = {R^{2}\over z^{2}} (-dt^2 +d\vec{x}^{2} +dz^2 )
+ R^{2}d\Omega_{5}^2~.
\ee
Making the static gauge choice $\sigma^{1}=t, \sigma^{2}=z$,
and restricting attention to
configurations with the string pointing along a particular $\bS^{5}$
direction,\footnote{The operator $\cO_{F^{2}}$
couples to the spherically symmetric mode of
the ten-dimensional dilaton, so we will focus attention on this
mode alone.  A string which is localized on the five-sphere
will excite
also all of the higher Kaluza-Klein harmonics, which are massive
fields on $AdS_{5}$. These excitations would give expectation values
to dual higher-dimension operators which have been identified in
\cite{gkpads,wittenholo,iwm}.}
the action reduces to
\be \label{string}
S_{F}= -{R^{2}\over 2\pi\ap} \int dt\,{dz\over z^{2}}
\sqrt{1-\p_{t}\vX^{2}+\p_{z}\vX^{2}
-\p_{t}\vX^{2}\p_{z}\vX^{2}
+\left(\p_{t}\vX\cdot\p_{z}\vX\right)^{2}},
\ee
where $\vX (z,t)$ denotes the position of the string in the
$\vx$ directions. The static solutions to\req{string}
can be taken to lie in the $z-x^{1}$
plane without loss of generality. They satisfy
\be \label{geodesic}
\p_{z}X_{s}=\pm{z^{2}\over\sqrt{z_{m}^{4}-z^{4}}}.
\ee
This equation describes a string lying along a geodesic which
starts and
ends at $z=0$ and reaches a maximum at $z=z_{m}$ (see
Fig.~\ref{bentstring}).
The two endpoints of the string are separated by a coordinate distance
\cite{reyee,juanwilson}
\be \label{separation}
L=z_{m}{(2\pi)^{3/2}\over\Gamma(1/4)^{2}}~.
\ee

Now consider small oscillations about the solution described
by\req{geodesic},
letting $\vX(z,t)=\vX_{s}(z)+\vY(z,t)$. For simplicity, we take
$\vY\perp\vX_{s}$. The linearized equation of motion for \vY\ is
\be \label{flucteom}
-\p_{t}^{2}\vY+\left[1-{z^{4}\over
z_{m}^{4}}\right]\p_{z}^{2}\vY-{2\over z} \p_{z}\vY =0~.
\ee
In the calculation to follow, we
will cut off $AdS_{5}$ by moving the boundary in to $z=z_{0}$ and
take $z_{0}\to 0$  at the end of the calculation. In order to solve
\req{flucteom}, we need boundary conditions for the left and right string
endpoints which we will impose in the form
$\vY_{L,R}(z_{0},t)=\vy_{L,R}(t)$.
The interpretation is straightforward: for a given $z_0$, the string is attempting
to describe a Higgsed gauge boson of very large mass ($\propto z_0^{-1}$)
transforming in the fundamental of the unbroken $SU(N)$ gauge group;
this massive object is an extrinsic degree of freedom from the point of
view of the $SU(N)$ gauge theory and has its own dynamics; this dynamics
is essentially that of a point particle and is thus described by a trajectory
function $\vy(t)$. For the moment, we will simply prescribe a trajectory,
but the Nambu-Goto action for the string in the $AdS_5$ geometry implies
an action for $\vy(t)$ which in turn implies an equation of motion for the
trajectory. We will not pursue this line of thought much further in this
paper, but it is interesting to note that the kinetic term in this
equation of
motion implies a quark mass that matches the static total energy of the quark/string.

Since the Nambu-Goto action\req{ng} depends on the background supergravity
fields, it is a source for them as well. In particular, it is a source
for the dilaton, a fact which is best displayed by writing the action
in terms of the Einstein metric $G^{E}_{MN}=e^{-\phi/2}G_{MN}$:
\be \label{nge}
S_{F}= -{1\over 2\pi\ap} \int dt\,dz\,e^{\phi/2} \sqrt{-g_{E}}.
\ee
The same metric rescaling in the bulk supergravity action yields
a dilaton kinetic term
\be \label{dilkin}
S_{S}=-{\Omega_{5}R^{5}\over 4\kappa^{2}}\int d^{5}x \,
\sqrt{-G_{E}}G_{E}^{mn}\p_{m}\phi\p_{n}\phi~.
\ee
Notice that the original ten-dimensional action has been dimensionally
reduced to $AdS_5$ over $\bS^{5}$: $\phi$ now denotes the projection of
the original ten-dimensional $\phi$ onto the constant $\bS^{5}$ spherical
harmonic and is a function on $AdS_5$ while $m,n$ are $AdS_{5}$ indices.
The combined action $S_{bulk}=S_{S}+S_{F}$
implies a linearized dilaton equation of motion
\be \label{dileom}
\p_{m}\left[\sqrt{-G_{E}}\,G_{E}^{mn}\p_{n}\phi\right]=J, \qquad
J(x)={2\kappa^{2}\over 4\pi\ap\Omega_{5}R^{5}}
     \sqrt{-g_{E}}\,\delta\left(\vx-\vX(z,t)\right)~.
\ee
This equation is solved by by Greens' function methods as
$\phi(x)=\int d^{5}x'\, D(x,x') J(x')$,
where
$D(x,x')$ is the retarded dilaton propagator \cite{dkk}.
The propagator is in fact only
a function of the invariant distance $v$, defined by
\be \label{v}
\cos v = 1- \frac{(t-t')^{2}-(\vx-\vx')^{2}-(z-z')^{2}}{2zz'}~.
\ee
Explicitly,
\be \label{prop}
D(v)=-{1\over 4\pi^{2}R^{3}\sin v}\,{d\over dv} \left[ {\cos 2v
\over\sin v} \theta(1-|\cos v|)\right]~.
\ee
This is a fairly complicated-looking propagator, but it is just the
dimensional reduction of the much simpler, completely algebraic
ten-dimensional $AdS_5\times \bS^{5}$ propagator
\be \label{algprop}
K \sim {(z z')^4 \over \left [(z \hat n - z' \hat n')^2 + (t-t')^2 +
(\vec x - \vec x')^2 \right ]^4 }
\ee
where the unit vectors $\hat n, \hat n'$ indicate position on $\bS^{5}$.

Having obtained $\phi(x)$ in the bulk, the GKPW recipe to extract the
expectation value is \cite{gkpads}
\be \label{trfsq}
\brac{\cO_{F^{2}}}=-{\delta S_{bulk}\over
\delta\phi}.
\ee
The expectation value would of course vanish in the gauge theory
vacuum sector. On the other hand, the string corresponds to the
sector of the gauge theory where a heavy quark has been inserted
in the vacuum. We do expect a non-zero expectation value of the
$\tr F^{2}$ operator in that sector and Eq.\req{trfsq} gives a
method for computing it. Carrying out similar steps with higher
$\bS^{5}$ harmonic modes of the dilaton would yield gauge theory
expectation values for operators of the type $\tr(F^2 X_I\ldots X_J)$,
where the $X_I$ are the scalar fields
of the $\cN=4$ gauge theory \cite{gkpads, wittenholo,iwm}.
These higher-dimension operators should give rise to a correspondingly
higher power law falloff with $|\vx|$, a result which should emerge naturally
from the structure of the Greens' functions for the higher $\bS^{5}$
harmonic modes of the dilaton.

Under $\phi\to\phi+\delta\phi$ the action varies only by a surface
term, because the configuration about which we vary is a solution
to the equation of motion:
\be \label{var}
\delta S_{bulk}={\Omega_{5}R^{8}\over 2\kappa^{2}}
 \left.\int dt d^{3}\vx\,\left({1\over z^{3}}\p_{z}\phi
 \delta\phi\right)\right|_{z=z_{0}}.
\ee
As a shorthand, it will be convenient to define a rescaled dilaton field
$\tphi=\Omega_{5}R^{8}\phi/ 2\kappa^{2}$. It follows from the foregoing
discussion that
\be \label{dilsol}
\tphi(x)=-{1\over 16\pi^{3}\ap}\int dt'\,dz'\,\sqrt{-g_{E}}
  {1\over\sin v}\,{d\over dv} \left[ {\cos 2v
\over\sin v} \theta(1-|\cos v|)\right],
\ee
and
\be \label{trfsq2}
\brac{\trFsq}=-
\left.\left({1\over z^{3}}\p_{z}\tphi \right)\right|_{z=z_{0}\to 0}.
\ee
Our task, then, is to calculate the dilaton field produced by various string
sources and to pick out the $O(z^4)$ term in its expansion near the
boundary
of $AdS_5$.

\section{Gauge Fields of an Oscillating Quark}
\label{straightsec}

We will examine first the especially simple
case $z_{m}\to\infty$, where the static solution describes
a straight string extending along the radial direction,
$\vX_{s}(z)=0$.  Eq.\req{flucteom} simplifies to
\be \label{flucteom2}
-\p_{t}^{2}\vY+\p_{z}^{2}\vY-{2\over z} \p_{z}\vY =0~.
\ee
This equation is
most easily solved via Fourier transformation.
The solution describing
purely outgoing waves is found to be
\be \label{straightfluct}
\vY(z,t)=\int d\omega\, e^{-i\omega(t-z+z_{0})}\left({1-i\omega z
\over 1-i\omega z_{0}}\right)\vy(\omega)~.
\ee
For simplicity, we specialize to harmonic boundary
data,\footnote{Henceforth it is
understood that one should take the real part of expressions
like this.}
$\vy(t)=\vec{A}\exp(-i\omega t)$.

The Nambu-Goto square root in\req{dilsol} can be expanded as
\be \label{straight}
\sqrt{-g_{E}}\simeq {R^{2}\over z^{2}}\left[1-{1\over 2}
               \left(\p_{t}\vY\right)^{2}+
                 {1\over 2}\left(\p_{z}\vY\right)^{2}\right].
\ee
Keeping only the leading term,\req{dilsol} reads
\be \label{dilstraightosc}
\tphi(\vx,z,t)=-\frac{R^{2}}{16\pi^{3}\ap}
     \int {dt'\over\sin v} \,{dz'\over z'^{2}}
    {d\over dv} \left[ {\cos 2v
      \over\sin v} \theta(1-|\cos v|)\right].
\ee
Next, change variables of integration $t'\to v$, using\req{v}, and
integrate by parts on $v$, to be left with
\bea \label{dilstraightosc2}
\tphi&=&\frac{R^{2}z}{16\pi^{3}\ap}
  \int{dz'\over z'}\,I, \nonumber \\
I&=& \int^{\pi}_{0} dv \,{\cos 2v \over\sin v} {d\over dv} \left[
   \frac{1}
   {\sqrt{z^{2}+z'^{2}+(\vx-\vY)^{2}-2zz'\cos v}}\right].
\eea
Now expand the square root in powers of $Y$. The leading
($Y$-independent) term gives rise to a static component of the
dilaton field, $\tphi_{s}(\vx,z)$. Its contribution to the gauge theory 
expectation value has been computed in \cite{dkk} and found to be
\be \label{expstraightstat}
\brac{\trFsq}_{s}={\sqrt{2}\over 32\pi^{2}}
 {\sqrt{g_{YM}^{2}N}\over{|\vx|^{4}}}~.
\ee
This is as expected for a point charge of magnitude proportional to
$(g_{YM}^{2}N)^{1/4}$, which is the effective strength of the
coupling as inferred from the quark-antiquark potential
\cite{reyee,juanwilson}.

The next term in expansion of\req{dilstraightosc2} in powers of $Y$,
the term linear in $Y$, gives the leading dynamical contribution to
$\trFsq$:
\bea \label{dilstraightosc2b}
\tphi_{(1)}&=&\frac{R^{2}z }{16\pi^{3}\ap}
  \int{dz'\over z'}\,I_{(1)}, \nonumber \\
I_{(1)}&=& \int^{\pi}_{0} dv \,{\cos 2v \over\sin v} {d\over dv} \left[
   \frac{\left(\vx\cdot\vec{A}\right)
          e^{-i\omega(t'-z')}\left(1-i\omega z' \right)}
   {\sqrt{z^{2}+z'^{2}+|\vx|^{2}-2zz'\cos v}}\right],
\eea
where $t'$ is understood to be a function of $v$,
$$
t'=t-\sqrt{z^{2}+z'^{2}+|\vx|^{2}-2zz'\cos v}.
$$
Since, according to\req{trfsq2}, we eventually only need the
$O(z^{4})$ terms in \tphi, we have set $z_{0}\to 0$
in\req{dilstraightosc2b}.

\begin{figure}[htb]
\centerline{\epsfxsize=9cm
\epsfbox{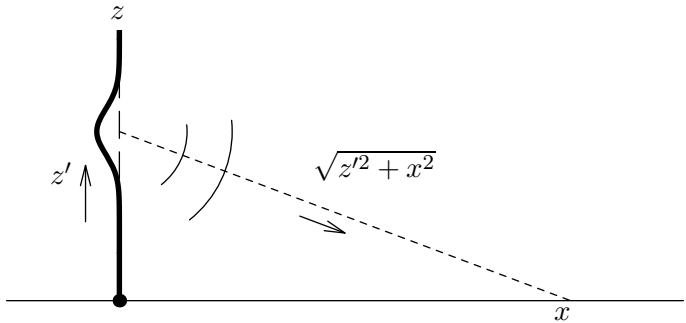}}
\caption{\small Any given point on the boundary receives radiation
from each point along the string. As a result, the gauge theory wave
is a superposition of components with all possible time delays. See
text for discussion.}
\label{signal}
 \begin{picture}(0,0)
 \put(104,26){\small $x$}
 \put(37,44){\small $z'$}
 \put(72,45){\small $\sqrt{z'^{2}+x^{2}}$}
 \put(45,66){\small $z$}
 \end{picture}
\end{figure}

If one expands the integrand of $I_{(1)}$ in powers of
$\eta=zz'\cos v/(z^{2}+z'^{2}+|\vx|^{2})$, the first non-vanishing term
is found to be $O(\eta^{3})$, and higher-order terms will not contribute
to\req{trfsq2}.  Keeping only the relevant terms one obtains
\bea \label{dilstraightosc3}
\tphi_{(1)}&=&-\frac{R^{2}\left(\vx\cdot\vec{A}\right)
                     z^{4}}{128\pi^{2}\ap}
  \int_{0}^{\infty}{dz'}\, z'^{2}(1-i\omega z')
  e^{-i\omega(t-\sqrt{z'^{2}+|\vec{x}|^{2}}-z')}
    f(\sqrt{z'^{2}+|\vx|^{2}}), \nonumber \\
f(u) &=& {i\omega^{3}\over u^{6}}-{12\omega^{2}\over u^{7}}
       -{57i\omega \over u^{8}}
         +{105\over u^{9}}~.
\eea
The bulk dilaton field is evidently a superposition of
waves radiated from each point along the string. The phase delay
$z'+\sqrt{z'^{2}+|\vx|^{2}}$ is simply the time needed for a
null signal to propagate up along the string to the point $z=z'$, and
then travel down diagonally to reach the boundary at $\vx$ (see
Fig.~\ref{signal}).

To understand this result from the viewpoint of the boundary theory,
it is advantageous to change the variable of integration to
$\zeta=\sqrt{1+z'^{2}/|\vx|^{2}}+z'/|\vx|$.
Using\req{dilstraightosc3} in\req{trfsq2} one
finds, after some integration by parts,
\bea \label{expstraightosc}
\brac{\trFsq}_{(1)}&=&
  \frac{\sqrt{2g_{YM}^{2}N}\left(\hat{x}\cdot\vec{A}\right)}
        {2\pi^{2}|\vec{x}|^{4}}\left\{
  \int_{1}^{\infty}{d\zeta}\, i\omega
  e^{-i\omega(t-\zeta|\vec{x}|)} \chi(\zeta)
     +{59\over 32|\vec{x}|}e^{-i\omega(t-|\vec{x}|)}\right\}\nonumber\\
\chi(\zeta)&=&\frac{210\zeta^{10}-258\zeta^{8}+267\zeta^{6}
        +69\zeta^{4}+55\zeta^{2}+1}
    {2\left(\zeta^{2}+1\right)^{7}}~.
\eea
The expectation value has been expressed solely in terms of gauge theory
quantities through use of the relation
$R^{2}/\ap=\sqrt{2g_{YM}^{2}N}$,
as must be possible for a proper gauge theory interpretation. It should
be noted that the dependence of the integrand on
$\omega|\vx|$ can be shifted from the phase factor to
the envelope function $\chi$ through an integration by parts.

Eq.\req{expstraightosc} displays the gauge theory disturbance
as a superposition of components propagating at speeds $v=1/\zeta$,
for all $1\le\zeta\le\infty$.
Notice that the weight factor $\chi\to 0$ as $\zeta\to\infty$,
so low velocity components are evidently suppressed.
It should be noted from\req{straightfluct}
that the string oscillations
actually become large (and our approximations fail)
for large $A\omega z'$, so the detailed shape of $\chi$
cannot be trusted at arbitrarily large $\zeta$.
Nonetheless, it is clear from the geometric setup (summarized in
Fig.~\ref{signal}) that the gauge theory wave should indeed
include components propagating at arbitrarily low velocities.

The above result implies in particular that even if the
charge is shaken abruptly to generate a sharply defined pulse on the
string, the SYM observer at $\vx$ will receive an infinitely broadened
pulse, only the leading edge of which travels at the speed of light. The
delayed signals presumably arise from rescattering of the original disturbance
from the static background\req{expstraightstat} by virtue of the
nonlinear
dynamics of the strongly-coupled gauge theory.
This rather complex sequence of events
would be difficult to unravel in the gauge theory, but the bulk-boundary
correspondence gives a precise and physically plausible account of it.
Note that the long time delays originate
from string
disturbances at large values of the bulk $AdS$ coordinate $z$, as would be
expected from the UV/IR connection proposed in \cite{susswi,pp}.

It is tempting to speak of\req{expstraightosc} as electromagnetic
`radiation', but the rapid falloff with $|\vx|$ indicates that
this would not be strictly correct.
We are looking at
a contribution to $\trFsq$ {\emph{linear}} in the displacement of the quark
and this can only arise from a cross-term between the static field and the
fluctuating field. Since the
static electric field is radial and the asymptotic
radiation gauge fields (the ones that fall off as $|\vx|^{-1}$) are
transverse, their
scalar product vanishes. Hence there is no unambiguous contribution of
electromagnetic
radiation to the $\trFsq$ expectation value: instead, we see evidence of
fluctuations
in the near-fields of the moving quarks, fields which do not transport
energy to infinity.
The unambiguous diagnostic for radiation would be the
demonstration of a net energy flux to spatial infinity in the gauge theory.
This could be done by
determining the expectation value of the gauge theory energy-momentum tensor,
which in the GKPW recipe is dual
to the bulk gravitational field produced by the fluctuating string.
It would be extremely
interesting to carry out this calculation explicitly, for
it is not at all obvious how (or even if) the $AdS$
description of waves in the boundary theory
incorporates energy conservation.
In this connection, we should also remark that our analysis neglects
the back-reaction on the string due to the supergravity fields.
We will return to these issues in Section \ref{implisec}.

Before closing this section, we note that our external charges are by
construction infinitely massive, and consequently immune to the SYM field
configuration they help to produce. It is possible to consider instead
sources with finite mass, represented by strings which
terminate not at the boundary, but on a solitary D3-brane placed at
$z_{b}>0$. In that case one is really studying an $SU(N+1)$
gauge theory, broken spontaneously to $SU(N)\times U(1)$ by a Higgs
vacuum expectation value $R^{2}/z_{b}\ap$ \cite{jthroat,dougtay}.

\section{Gauge Fields of Heavy Quark Mesons}
\label{bentsec}

We now extend the analysis to the general case $z_{m}<\infty$, where the
string bends along the geodesic\req{geodesic}. Both of
its endpoints reach the boundary, so this configuration
describes a quark-antiquark pair (see Fig.~\ref{bentstring}).
Notice that now the parametrization $\vX(z,t)$
has the disadvantage of being double-valued: for each value of $z$
there are in fact two points on the string, one on the left
and one on the right. When necessary, we will account for this by
means of a discrete subindex: $\vX_{L,R}(z,t)$. The need for this
awkward notation is compensated by the simple
form of the differential equation\req{flucteom}.

The expansion of the Nambu-Goto integrand now yields
\be \label{bent}
\sqrt{-g_{E}}\simeq {R^{2}\over z^{2}}\left\{\Delta
+{1\over 2\Delta}
  \left[\left(\p_{z}\vY\right)^{2}-\Delta^{2}
  \left(\p_{t}\vY\right)^{2}\right]\right\},
\qquad \Delta={z_{m}^{2}\over\sqrt{z_{m}^{4}-z^{4}}}.
\ee
The dilaton\req{dilsol}
is again a sum of static and fluctuating components.

It is interesting to
determine the gauge field profile due to the static
bent string.
Inserting the first term of\req{bent} into\req{dilsol}, changing
variables $t'\to v$, and integrating by parts with respect to $v$ one
obtains
\bea \label{dilbentstat}
\tphi_{s}(\vx,z)&=&\frac{R^{2}z_{m}^{2}z}{16\pi^{3}\ap}
  \int{dz'\over z'\sqrt{z_{m}^{4}-z'^{4}}}\,I, \nonumber \\
I&=& \int^{\pi}_{0} dv \,{\cos 2v \over\sin v} {d\over dv} \left[
   {1\over\sqrt{z^{2}+z'^{2}+(\vx-\vX(z'))^{2}-2zz'\cos v}}\right]~.
\eea
Next, expand the integrand of $I$ in powers of
$2zz'\cos v/[z^{2}+z'^{2}+(\vx-\vX')^{2}]$,
and retain only the leading order term, to find
\be \label{i}
I= -{15\pi(z z')^{3}\over 8
\left[z^{2}+z'^{2}+\left(\vx-\vX(z')\right)^{2}\right]^{7/2}}~.
\ee
Use of this in\req{dilbentstat} leads to
\bea \label{dilbentstat2}
\tphi_{s}&=&-\frac{15R^{2}z_{m}^{2}z^{4}}{128\pi^{2}\ap}
 \int_{0}^{z_{m}}{dz' z'^{2}\over \sqrt{z_{m}^{4}-z'^{4}}}
  \nonumber \\
{}&{}&\times \left\{{1\over\left[z'^{2}+\left(\vx-\vX_{L}(z')
     \right)^{2}\right]^{7/2}}
 +{1\over\left[z'^{2}+\left(\vx-\vX_{R}(z')
     \right)^{2}\right]^{7/2}}\right\},
\eea
where we have explicitly indicated the contribution
from both halves of the string.
It is convenient to place the center of the string at the origin,
$\vX(z_{m})=0$, so that $X_{L}(z)=-X_{R}(z)$,
as depicted in Fig.~\ref{bentstring}.

We wish to extract the leading
term in\req{dilbentstat2} for $|\vx|\gg L$, which by\req{separation}
implies that $|\vx|\gg z_{m}$ as well.
We find
\be \label{dilbentstat3}
\tphi_{s}=-\frac{15R^{2}z_{m}^{2}z^{4}L}
                 {128\pi^{2}\ap|\vx|^{7}}~.
\ee
The SYM expectation value then follows from\req{trfsq2}.
We can express the result in terms of quantities in the boundary
theory,
using\req{separation} and $R^{2}/\ap=\sqrt{2g_{YM}^{2}N}$:
\be \label{expbentstat}
\brac{\trFsq}_{s}={15\Gamma(1/4)^{4}
\sqrt{2}\over 8 (2\pi)^{5}}
 {L^{3}\sqrt{g_{YM}^{2}N}\over{|\vx|^{7}}}~.
\ee
Notice the peculiar dependence on $L$ and $|\vx|$ and the fact
that the result is isotropic. This is not what one would expect
from a static electric dipole  field in a linear gauge theory, but there
is nothing obviously inconsistent about it for strongly coupled $\cN=4$ SYM.
The above result must be regarded as a prediction of the bulk-boundary
correspondence for which we have at present no independent test.

We now examine the contribution from the fluctuating part.
Equation\req{flucteom} can be solved by Fourier transformation.
The general solution is
\bea \label{bentfluct}
\vY_{L,R}(z,t)&=&\int d\omega\, \sqrt{1+\omega^{2}z^{2}}
\left\{\vec{A}(\omega)e^{-i\omega(t-\cZ_{L,R})}
      +\vec{B}(\omega)e^{-i\omega(t+\cZ_{L,R})}\right\}, \nonumber\\
\cZ_{L}(z,\omega)&=&\sqrt{(\omega z_{m})^{4}-1}\int_{z_{0}}^{z}
    \frac{(s/z_{m})^{2}ds}{\left(1+\omega^{2}s^{2}\right)
      \sqrt{1-(s/z_{m})^{4}}}, \\
\cZ_{R}(z,\omega)&=&\cZ_{L}(z_{m},\omega)
   +\sqrt{(\omega z_{m})^{4}-1}\int_{z}^{z_{m}}
    \frac{(s/z_{m})^{2}ds}{\left(1+\omega^{2}s^{2}\right)
      \sqrt{1-(s/z_{m})^{4}}}~. \nonumber
\eea
Notice that component waves with
$\omega z_{m}<1$ are exponentially damped, reflecting a
frequency cutoff imposed by the finite size of the
string.
The oscillations on the left and right halves of the
string are related by the requirement that the solution be smooth at
the midpoint, $z=z_{m}$. The coefficients $\vec{A}$ and $\vec{B}$
are determined
by enforcing boundary conditions at the string endpoints,
$\vY_{L,R}(z_{0},t)=\vy_{L,R}(t)$:
\be \label{coeffs}
\vec{A}(\omega)=\frac{\vy_{L}(\omega)-\Phi\vy_{R}(\omega)}
  {1-\Phi^{2}}, \quad
\vec{B}(\omega)=\frac{\vy_{L}(\omega)-\Phi^{*}\vy_{R}(\omega)}
{1-\Phi^{*2}}, \quad
\Phi(\omega)=e^{i2\omega\cZ_{L}(z_{m},\omega)}~.
\ee

For any given choice of boundary conditions,
the string endpoints trace out definite Wilson lines $\vy_{L,R}(t)$
in the gauge theory.
The solution\req{bentfluct}
can be used in\req{dilsol} and then\req{trfsq2} to determine the
corresponding SYM expectation value.
Rather than working out the details of such a calculation,
which would not be particularly enlightening,
we will point out some interesting general
features of the resulting field configurations.

First, it is evident that the SYM waves  display a
phase delay analogous to the one found for the straight string,
although the details are different. To understand this in some
detail,
imagine that at $t=0$ a pulse is sent along the string by shaking
its left end, which we now take to be located at $\vx=0$.
The induced metric on the bent string is
\be \label{g}
g_{ab}d\sigma^{a}d\sigma^{b}=
    {R^{2}\over z^{2}} \left[-dt^2
    +{dz^2\over 1-(z/z_{m})^{4}}\right],
\ee
so the pulse, following
a null trajectory, takes a time
\be \label{time}
\Delta t_{1}(z')=\int_{0}^{z'}
     {dz\over\sqrt{1-(z/z_{m})^{4}}}\quad \mbox{or}\quad
    \left(\int_{0}^{z_{m}}+ \int_{z'}^{z_{m}}\right)
     {dz\over\sqrt{1-(z/z_{m})^{4}}}
\ee
to reach the point $z'$ on the left or right half of the string.
In particular, it requires a time \cite{bakrey}
\be \label{delay}
T=z_{m}{\Gamma(1/4)^{2}\over2\sqrt{2\pi}}
\ee
to traverse the entire string
and arrive at the right endpoint, where
for the time being we assume that it is completely absorbed.

\begin{figure}[htb]
\centerline{\epsfxsize=10.2cm
\epsfbox{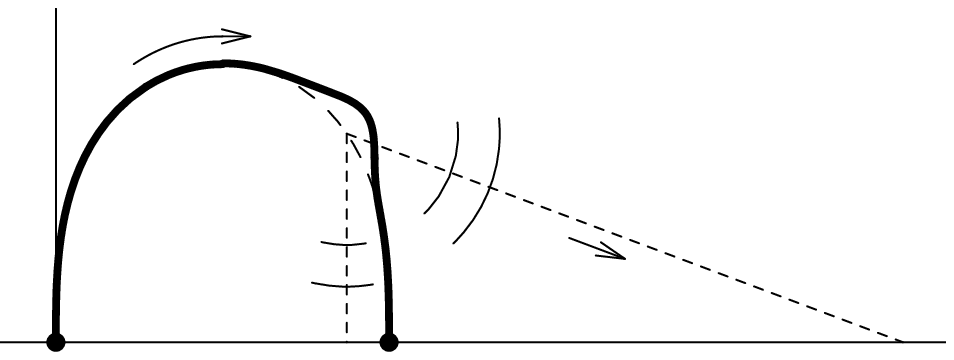}}
\caption{\small Any given point on the boundary receives radiation from
each point along the `bent' string, with a corresponding time delay.
Notice that the boundary projection of wavefronts given off at a
point $z'$ on the string yields a spherical wave which appears to
originate from $\vX_{s}(z')$. See text for discussion.}
\label{signal2}
 \begin{picture}(0,0)
 \put(120,30){\small $x$}
 \put(58,53){\small $z'$}
 \put(40,69){\small $\Delta t_{1}$}
 \put(90,50){\small $\Delta t_{2}$}
 \put(30,71){\small $z$}
 \put(55,30){\small $X_{s}(z')$}
 \end{picture}
\end{figure}

As seen in Fig.~\ref{signal2},
a dilaton wave travels from $z'$ down to a boundary point $\vx$
in an additional time
$\Delta t_{2}(z',\vx)=\sqrt{z'^{2}+\left(\vx-\vX_{s}(z')\right)^{2}}$.
As a result,
the radiation arriving at $\vx$ has a component
with phase lag $\Delta t_{1}(z')+\Delta t_{2}(z',\vx)$
for each point $z'$ on the string. The net effect is that
the SYM observer detects a significantly broadened
pulse, whose leading and trailing edges arrive at
times $t_{f}=|\vx|$ and $t_{b}=T+|\vx-L\hat{x}_{1}|$,
respectively.

The situation is thus similar to the one encoded
in\req{expstraightosc}, in that an oscillating source would
ultimately give rise to a superposition of gauge theory
waves traveling at  different speeds $v\le 1$. A complicating
feature of this situation as compared to the case of an isolated
quark is that, because the string now extends only a finite
distance into $AdS$ space, a disturbance on the string can only
propagate for a
finite time before running into the boundary. As has been observed
by others \cite{bakrey}, the time\req{delay} for a disturbance
to propagate from one end of the string to the other corresponds,
from the gauge theory point of view, to a subluminal mean speed of
propagation of influence $v=L/T\simeq 0.457$. Note, however, that
this is not the generic speed of propagation of disturbances in
the gauge theory: the whole point of our analysis was to show that
disturbances in the expectation value of $\trFsq$ propagate away
from their source in the quark-antiquark system in a conventionally
causal fashion: the leading signal arrives on a direct path at the
speed of light, followed by indirect signals that arrive later.

As an aside, we remark that disturbances propagating on the strings
have bizarre features from the point of view of the boundary gauge
theory. For instance, triple-string configurations
(describing for example a quark-monopole-dyon system \cite{danpoly})
can be arranged where a signal originating from one charge would give
rise to a disturbance which would run along the strings and arrive
first at the more distant (from the boundary theory perspective) of the two
other charges \cite{gt}. This would not violate causality, strictly
understood, but is certainly strange. To repeat,
the key point is that an oscillation on the string does not translate
{\emph{directly}} into a wave in the boundary theory. The correct
prescription brings the bulk supergravity fields into play,
and unequivocally predicts causal SYM propagation,
with propagation velocities up to
the speed of light.

\section{Implications for Gauge Theory Dynamics}
\label{implisec}

In this paper, we have been exploring a picture,
derived from Maldacena's AdS/CFT duality conjecture, of
the generation and propagation of disturbances in the $D=3+1$
$\cN=4$ $SU(N)$ super-Yang-Mills (SYM) gauge theory. In this
picture, external sources are described by type IIB strings
running from the boundary into the bulk of $AdS$ space;
fluctuations in the position of the external sources generate
waves on the strings; the string waves generate propagating
disturbances in the  supergravity fields in $AdS$ space; finally,
the fluctuating boundary values of these fields are converted,
via the GKPW recipe, into fluctuating expectation values of
operators in the  gauge theory. Throughout the discussion, we have
assumed that the  string disturbances propagate according to simple
Nambu-Goto dynamics and have treated them as known linear sources
for the supergravity fields. In particular, we have not worried
about back-reaction of the supergravity fields on disturbances
propagating on the string. On the face of it, this seems reasonable
because Maldacena's conjecture includes taking the limit of weak
supergravity coupling. On the other hand, as we will now discuss,
this collection of assumptions leads to some surprising, perhaps
paradoxical, conclusions about the behavior of the gauge theory
that are worth pointing out.

A somewhat
perplexing feature of the time-dependent field that
can be gleaned from Fig.~\ref{signal2} is that dilaton wavefronts
emitted from a point $z'$ on the string describing a quark-antiquark
system, give rise to spherical waves in the gauge theory which seem to
emanate from neither the quark nor the antiquark, but from the point
$\vX_{s}(z')$ on the line between them. Imagine an observer situated
halfway
between the quark and antiquark: if the quark is shaken to produce a
pulse
on the string, the observer first sees disturbances coming first from
the direction of the quark and then (after a time $T/2$)
from the opposite direction! Though odd, this feature is in principle
consistent with the non-linear character of strongly-coupled SYM: the
external
sources give rise to
propagating disturbances, which propagate through and cause
to reradiate, the background gauge field configuration originally set
up by the source. This sort of thing would happen in any strongly-coupled
theory; what is surprising is the geometrical structure that is inherited
from the $AdS$ string.

A more profound set of issues
arises from the fact that a disturbance travels from one
end of the quark-antiquark string to the other in a finite time\req{delay},
forcing us to consider how the string disturbance reflects from the
boundary if we wish to account for radiation generated at later times.
Since the external sources can be taken to be as massive as we like
(by letting $z_0\to 0$), it seems reasonable to assume that the fluctuating
string should be subject to fixed or Dirichlet boundary
conditions\footnote{The boundary conditions appropriate
for Wilson loops in the $AdS$/CFT correspondence have been discussed
in \cite{drugroguri}.}
which
reflect any incident disturbance back onto the string (with a change of sign).
This would mean that a disturbance, however it was initially generated,
would simply reflect back and forth between the quark and antiquark ends
of the string without ever dying away. More precisely, the linearized
string
would have eigenstates of oscillation at a discrete set of frequencies
$\omega_{n}$ running from some lower cutoff on up to infinity. In a WKB
approximation, these frequencies would be determined by the requirement that
the phase factor $\Phi$ in\req{bentfluct} is real, i.e.
\be \label{harmonics}
\omega_{n}z_{m}\sqrt{(\omega_{n}z_{m})^{4}-1}\int_{0}^{1}
   {d\sigma\over\left[ 1+(\omega_{n}z_{m})^{2}\sigma^{2}\right]
    \sqrt{1-\sigma^{2}}} = {n \pi\over 2}~.
\ee
These oscillations must represent excited states of the dipole
field, with a mass gap between states scaled by the dipole separation $L$.
These states are quite analogous to the infinite tower of mesons found in
the large-$N_c$ limit of ordinary QCD (where the mass gap is set by
the confinement scale). On the other hand, it is quite surprising to
imagine finding an analogous set of states in a non-confining
conformal gauge theory!

At this point we are led back to the questions, first raised in the
discussion of the isolated quark in Section \ref{straightsec},
of radiation, energy conservation and
back-reaction. It is important to realize that
a complete treatment of the production of time-varying
supergravity fields by a disturbance on the string must include
back-reaction on the string disturbance. To the extent that the  bulk field
includes a net energy flux away from the string, the back-reaction
should cause the disturbance to damp as it propagates. This process
is essential to energy conservation in the supergravity picture.
Let us now try to understand how this translates
into energy conservation in the gauge theory--- we will be led to
a paradox.

Focus attention again on the infinite tower of excited states
of the SYM dipole system, and consider the following question:
are these excited states stable? To answer this question,
we will pursue two possible lines of argument.
On the one hand, within the supergravity framework
we know that, once we take back-reaction into account, the
resonances will have finite widths and the notion of resonance will
only make sense if there is a limit in which the width becomes small
compared to the gap between successive states. The rate at which string
disturbances radiate is set by $g_s$, so if we take the usual $AdS$/CFT
limit $g_s\to 0$ (with $g_{s}N$ fixed), the string will
not radiate, and the excitations will be completely stable. This can
be seen explicitly in\req{dileom}: the source term in the dilaton
equation of motion vanishes as $g_s\to 0$ with $g_{s}N$ fixed.
We are thus led to conclude that
in the $N\to\infty$ limit (with the `t~Hooft coupling $g_{YM}^{2}N$ fixed)
there exists in the dual gauge theory
an infinite tower of stable (i.e., non-radiating) excited
states of the gauge field set up by an infinitely massive
quark-antiquark dipole.
As we have already pointed out, this would be
analogous to what happens in conventional QCD in the large-$N$ limit:
in the leading approximation, there is a tower of stable states in
every sector of the theory (meson, baryon, quarkonium, $\dots$); beyond
leading order, these states acquire finite widths proportional to some
power
of $1/N$. It would be most remarkable if the same structure of states
survived the passage from confining QCD to non-confining $\cN=4$ SYM
(with
the confinement scale replaced by a variable geometric scale set by the
`size' of the configuration).

On the other hand, the central point of this paper is that
the GKPW recipe translates a disturbance propagating on the
string into waves in the gauge theory.
At the end of the calculations
one obtains SYM expectation values (Eq.\req{expstraightosc}, for
instance) which depend on $g_{YM}$
only through the `t~Hooft coupling $g_{YM}^{2}N$,
and consequently {\emph{do not vanish}}\footnote{The reason for this
can be seen in\req{trfsq2}:
the gauge theory expectation value is extracted directly not from
the dilaton $\phi$ (which vanishes as $\gs\to 0$), but from the
rescaled field $\tphi\sim \phi/g_{s}^{2}$.}
when $\gs\to 0$.
Now, as discussed in Section \ref{straightsec},
the result obtained in\req{expstraightosc} is a near-field
contribution (it involves time-dependent fields which depend on the
velocity, but not the acceleration, of the sources), and so
does not unambigously indicate the presence of SYM radiation.
Nonetheless, given that the ten-dimensional
static, near-, and radiation fields all come in at the same order in
\gs\ (they differ only by their dependence on $|\vx|$),  it is
natural to expect that a computation of the energy-momentum tensor
would reveal a net energy flux away from the external sources,
signaling the presence of true radiation. On the face of it, this
seems to apply just as much to the solitary oscillating quark
as to the quark-antiquark excited states.

We have thus been led to a paradox:
if the gauge theory is to conserve energy, a radiating dipole field
cannot possibly be stable. To restate the problem in slightly
different words, imagine that the quark in the quark-antiquark
system is shaken abruptly to produce a pulse running along the string,
and the external charges are held fixed at all other times. In this
process, a definite amount of energy is added to the system.
In the $\gs\to 0$ limit, the pulse on the string will not decay, and so
it
will endlessly travel back and forth between the quark and antiquark.
Through the mechanism analysed in detail in this paper,
this disturbance will give rise to time-dependent SYM fields which
remain finite as $\gs\to 0$. If these fields include true radiation
(as seems reasonable to expect), they continuously carry energy away
from the dipole system, violating energy conservation in the gauge
theory. We should remark that, even though the paradox is most evident
in the context of the quark-antiquark system,
the question of energy conservation must also be addressed in the
case of the solitary quark. In that instance, the existence of SYM
radiation
would not in itself be paradoxical, but it is certainly far from obvious
that the infinitely broadened pulse which propagates
in the gauge theory after the external charge is shaken properly
incorporates energy conservation.

What are we to make of this?
In a sense, it is not surprising that we have encountered a problem:
given the holographic character of the $AdS$/CFT correspondence, the
interplay between energy conservation in the bulk and on the boundary
is bound to be a delicate issue.
Notice that the problem would disappear if our assumption regarding
the presence of radiation turned out to be erroneous.
Since we have not seen direct evidence for the
existence of gauge theory radiation in the $\gs\to 0$ limit, we must
bear in mind the possibility that the explicit determination of the
energy-momentum tensor will show that there is no net energy flux
away from the dipole system. This would undoubtedly be a surprising
result.
We will leave for future work
the more thorough analysis required
to reach a definitive conclusion on this important issue.

\section{Gauge Fields of Heavy Quark Baryons}
\label{barsec}

In the preceding two sections, we studied the gauge fields of a color-neutral
heavy quark-antiquark pair. Among other interesting things, we found
in the static case that the $\trFsq$ operator expectation value falls
off with distance like $|\vx|^{-7}$ (as compared to the $|\vx|^{-4}$ falloff
of the same quantity around an isolated color fundamental quark). To assess
how general this result is, we will now study the state of the gauge
field around a color-neutral collection of $N$ quarks:
the baryon of this gauge theory.

A baryon in $\cN=4$ SYM is dual to a fivebrane on which $N$
fundamental strings terminate \cite{wittenbaryon,groguri}. The precise
description of this system was found in
Ref.~\cite{cgs} (see also \cite{imamura}) through a study of
the fivebrane worldvolume action. In this approach
the strings are faithfully represented by a specific
deformation of the flux-carrying fivebrane, in accord with the
Born-Infeld string philosophy \cite{calmal,gibbons}.
The explicit fivebrane embedding that corresponds to a baryon was
found to be \cite{cgs}
\be \label{adssol}
r(\theta)=\frac{r_{0}}{\sin\theta}
          \left[\frac{3}{2}\left(\theta-\sin\theta \cos\theta\right)
          \right]^{1/3},
\ee
where $r=R^{2}/z$, $\theta$ is the $\bS^{5}$ polar angle, and
$r_{0}=r(\theta=0)$ is a modulus of the configuration. Since
the fivebrane is just as much a source of the dilaton as is the
string, we may use the logic of the earlier part of the paper to
infer the gauge theory expectation value  of $\trFsq$ in the presence
of a  baryonic collection of heavy quarks. The interesting question is
whether this approach yields the same scaling with $N$ and $|\vx|$ as
would the description of the baryon as a collection of quark strings.

Following \cite{cgs},
the fivebrane action for an embedding of the above type
in the presence of a nontrivial dilaton field can be seen
to read (in the Einstein frame)
\be \label{actionagain}
S_{D5}= T_5 \Omega_{4}R^{4}\int dt d\theta \sin^4\theta \{ -
  \sqrt{e^{\phi}\left[r^2+r^{\prime 2}\right]
  -E^2}  +4 A_0 \},
\ee
from which $E=F_{0\theta}$ may be eliminated in favor
of the displacement field
\be
D={\sin^4\theta E\over \sqrt{r^2+{r^\prime}^2-E^2}},
\ee
which is known explicitly as a function of $\theta$:
\be
D(\theta) = {3\over 2}\left(\sin\theta\cos\theta-\theta\right)
           +\sin^{3}\theta\cos\theta~.
\ee
After this replacement, Eq.\req{actionagain} implies a
linearized dilaton source term which can be written in the form
\be \label{dilsource2}
S_{D5\phi}= -{N\over 3\pi^{2}\ap}\int dt\,d\theta\,
\phi\sqrt{r^2+r^{\prime 2}}\sqrt{D^2+\sin^8\theta},
\ee
where we have made use of the relation $T_5 \Omega_{4}R^4=N/3\pi^{2}\ap$.

The embeddings of interest satisfy a BPS condition
\cite{imamura,cgs,gomis},
which can be used to eliminate $r^{\prime}$ in favor of $r$, yielding
\be \label{dilsource3}
S_{D5\phi}= -{N\over 3\pi^{2}\ap}\int dt\,d\theta\,r(\theta)
\left({D^2+\sin^8\theta \over
\sin^{4}\theta\cos\theta-D\sin\theta}\right) \phi
=-{N\over 3\pi^{2}\ap}\int dt\,dz\,f(z)\phi.
\ee
To make contact with the discussion of the present paper,
in the second step we have reparametrized the
fivebrane by the Poincar\'e radial coordinate $z=R^{2}/r$,
implicitly defining the function $f(z)$.

\begin{figure}[htb]
\centerline{\epsfxsize=7cm
\epsfbox{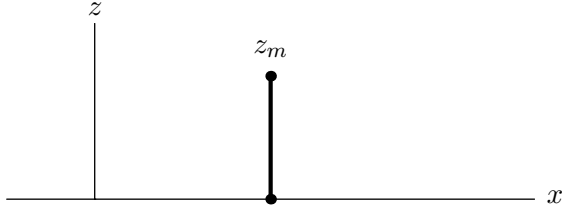}}
\label{bar}
  \begin{picture}(0,0)
  \put(52,34){\small $z$}
  \put(74,29){\small $z_{m}$}
  \put(113,9){\small $x$}
  \end{picture}
\caption{\small From the $AdS_{5}$ perspective, the baryon is a `fat
string' extending from the boundary up to $z=z_{m}$. At each value
of $z$, the fivebrane lies at a different polar angle on the
five-sphere.}
\end{figure}

It is important to note at this point that, unlike the string
configurations discussed in previous sections, which point along
a fixed direction on the five-sphere, the fivebrane has a
non-trivial $\bS^{5}$ dependence. At each value of $z$
it lies at a different polar angle $\theta=\theta(z)$
determined by\req{adssol}, and it is wrapped isotropically over
the remaining $\bS^{4}$.  The operator
$\trFsq$ couples to the massless Kaluza-Klein mode of the
dilaton, so the source term given by\req{dilsource3}
must be projected onto its spherically symmetric component.
In the case of the strings discussed in the preceding sections,
this projection would simply multiply the
source term by a numerical constant.
For the fivebrane, however, it
introduces an additional factor of $\sin^{4}\theta(z)$.
The resulting source for the massless $AdS_{5}$ dilaton is
\be \label{dilsource4}
J(x)={2\kappa^{2}\over 3\pi^{2}\ap\Omega_{5}R^{5}}
     f(z)\sin^{4}\theta(z)\,\delta\left(\vx\right)~.
\ee
It follows that the (rescaled) dilaton field is now given by
\be \label{newdilsol}
\tphi(x)=-{N\over 12\pi^{4}\ap}\int dt'\,dz'\,f(z')
  {\sin^{4}\theta(z')\over\sin v}\,{d\over dv}
  \left[ {\cos 2v \over\sin v} \theta(1-|\cos v|)\right]
\ee
(where the invariant distance $v$ is given in \req{v})
Through a familiar set of steps, one can extract the leading behavior
of $\tphi$ in the neighborhood of the $z=0$ boundary of $AdS_5$:
\be \label{dilfivestat}
{\tphi=-\frac{5 N z^{4}}{4(2\pi)^{3}\ap}
 \int_{0}^{z_{m}}dz'\,
 {z'^{4}f(z') \sin^{4}\theta(z')
  \over\left[z'^{2}+|\vx|^{2}\right]^{7/2}},}
\ee
where $z_{m}=R^{2}/r_{0}$ is the maximum value of $z$ to which
the fivebrane extends.

To obtain information
from\req{dilfivestat} it is convenient to return to the initial
angular parametrization:
\be \label{dilfivestat2}
\tphi=-\frac{5 N R^{8}z^{4}}{4(2\pi)^{3}\ap}
 \int_{0}^{\pi}
 \frac{\sin^{4}\theta \,d\theta}
 {r(\theta)^{3}\left[{R^{4}\over r(\theta)^{2}}
 +|\vx|^{2}\right]^{7/2}}
 \left({D^2+\sin^8\theta \over
\sin^{4}\theta\cos\theta-D\sin\theta}\right),
\ee
where the embedding $r(\theta)$ is given by\req{adssol}.
The complete field
profile of the baryon then follows from\req{trfsq2}.
{}From the way
$R^{4}/r^{2}=z'^{2}$ appears in the denominator
of\req{dilfivestat} and\req{dilfivestat2} it is clear that
the dilaton field (and consequently the SYM field profile)
will have qualitatively
different behavior in the regions $|\vx|>z_{m}$ and $|\vx|<z_{m}$,
so the modulus $z_{m}$ in fact determines
the `size' of the baryon, as expected from the UV/IR connection
\cite{susswi,pp}
(see e.g. the discussion in \cite{cgst}).

For $|\vx|\gg z_{m}$, the leading term in\req{dilfivestat2} is
\be \label{dilfivestat3}
\tphi=-\frac{5 N R^{2}z_{m}^{3}z^{4}}{9(2\pi)^{3}\ap|\vx|^{7}}
 \int_{0}^{\pi} d\theta
 {\sin^{6}\theta\over\left[\theta-\sin\theta\cos\theta\right]^{2}}
 (D^2+\sin^8\theta)~.
\ee
Letting $c\simeq 2.40$ denote the result of the angular integration and
employing\req{trfsq2}, we find that the $\trFsq$ expectation
value at large distance from the baryon is
\be \label{expfivestat}
\brac{\trFsq}={5c\sqrt{2}\over 18\pi^{3}}
 {z_{m}^{3}N \sqrt{g_{YM}^{2}N}\over{|\vx|^{7}}}~.
\ee
Notice that the dependence on $|\vx|$ and
the scale size of the configuration ($z_{m}$) is exactly the
same as that found for the `meson', Eq.\req{expbentstat}.
This is probably a generic feature of color-neutral objects
in the $\cN=4$ SYM gauge theory. From the string theory
perspective the common origin of this behavior is clear: unlike the
quark, the meson and the baryon are represented by brane objects
which do not extend all the way to the horizon at $z=\infty$.

A significant difference between\req{expbentstat}
and\req{expfivestat} is that the latter
includes an additional power of $N$. This is precisely as it should
be,\footnote{We thank Igor Klebanov for a discussion on this point.}
since $\tr F^{2}/4 g_{YM}^{2}$ should scale with $N$ in the same way as
the energy-momentum tensor: at fixed $g_{YM}^{2}N$ it should be
$O(1)$ for a meson, and $O(N)$ for an $SU(N)$ baryon
\cite{wittenbaryon}.

\section{Conclusions}

We have examined the correspondence between external
charges in $\cN=4$ SYM and strings in $AdS$ space.
Our principal focus was the connection between string oscillations and
gauge theory waves. Specifically, by studying the bulk radiation
given off by an undulating string, we determined the
time-dependent fields produced by an oscillating quark or a
quark-antiquark pair in the strongly-coupled theory. The picture
that emerges is one in which the waves are in fact
generated not only by the external sources, but also by the non-linear
medium supplied by the static background field of the same sources. 
This is in agreement with our qualitative expectations for
strongly-coupled non-Abelian gauge theory. The same considerations also
suggest the existence of an infinite tower of excitations 
in the quark-antiquark system in the extreme Maldacena limit.
The status of these excitations is uncertain, pending the resolution
of some puzzles regarding energy conservation in the $AdS$ description 
of the SYM theory, a subject to which we hope to return.

As a side-result, we have determined the static fields produced by a
quark-antiquark pair and also by the D-brane representative of the baryon. 
Both color-neutral systems were found to display the same long-distance 
behavior, and to have operator expectation values which fall off more rapidly 
with distance than those of the isolated quark.

Our results provide yet another
example of the remarkable way in which
the bulk-boundary correspondence manages to relate
intricate aspects of the
dynamics of strongly-coupled gauge theories
to properties of string theory in $AdS$ space.
At the same time, we have stressed the
need for further work to unravel the precise way
in which SYM energy conservation
manifests itself in the dual holographic description.

\section*{Acknowledgements}

We are grateful to Igor
Klebanov for helpful discussions.
AG would also like to thank Shiraz Minwalla,
{\O}yvind~Tafjord, and Mark~Van~Raamsdonk for useful conversations.
This work was supported in part by US Department of Energy
grant DE-FG02-91ER40671 and by National Science Foundation grant
PHY98-02484. AG is additionally supported by the National Science and
Technology Council of Mexico (CONACYT).

\end{document}